\documentclass[prb,twocolumn,showpacs,preprintnumbers,amsmath,amssymb]{revtex4}

\usepackage{graphicx}
\graphicspath{{fig/}}
\usepackage{dcolumn}
\usepackage{bm}
\usepackage{amsmath,amssymb}
\usepackage[T1]{fontenc}
\usepackage{textcomp}
\usepackage{color}

\begin{document}

\title{Wiedemann-Franz law for magnon transport
}

\author{Kouki Nakata,$^{1}$ Pascal Simon,$^2$ and Daniel Loss$^{1}$}

\affiliation{$^1$Department of Physics, University of Basel,   Klingelbergstrasse 82, CH-4056 Basel, Switzerland   \\
$^2$Laboratoire de Physique des Solides, CNRS UMR-8502, Universit$\acute{e} $ Paris Sud, 91405 Orsay Cedex, France  
}

\date{\today}

\begin{abstract}
One of the main goals of spintronics is to improve 
transport of information carriers and to achieve new functionalities with ultra-low dissipation. A most promising strategy for this holy grail
is to use pure magnon currents created and transported in insulating magnets,  in the complete absence
 of any  conducting metallic elements. Here we propose a realistic solution to this fundamental challenge by
analyzing  magnon and heat transport in insulating ferromagnetic  junctions. We calculate 
 all transport coefficients for magnon transport and establish Onsager relations between them. We theoretically discover that 
 magnon transport in  junctions has a universal  behavior, {\it i.e.} is independent of material parameters, and establish a magnon analog of the celebrated Wiedemann-Franz law which governs charge transport at low temperatures. We calculate the Seebeck and Peltier coefficients which are crucial quantities for spin caloritronics and demonstrate that they assume universal values in the low temperature limit. Finally,
we show that our predictions are within experimental reach with current device and measurement technologies.

\end{abstract}


\maketitle

{\section{Introduction}}
The study of thermoelectric properties of materials started more than two centuries ago and has been receiving recently  enormous attention motivated by the search of low-power consumption technologies. One of the key ingredients in thermoelectrics is the  Seebeck effect which refers to the generation of a voltage across the system under study  when a thermal gradient is externally applied to it. Conversely, the Peltier effect is traditionally associated with the generation of a heat flow when an electric current is driven across the  system. Despite the fact that most studies focus on electric conductors, these fundamental concepts also extend to spin degrees of freedom and are at the heart of the emerging field of spin caloritronics which  focuses on the interaction of spins with heat currents \cite{spincal,spincalreview,caloritronics,Sinova}.

\begin{figure}[h]
\begin{center}
\includegraphics[width=7.8cm,clip]{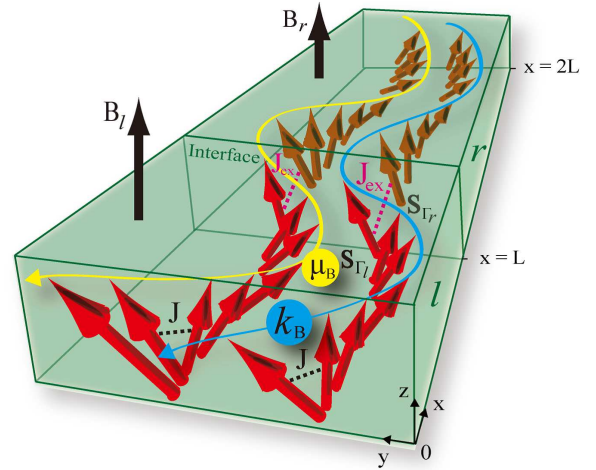}
\caption{(Color online)
Schematic representation of the ferromagnetic insulating junction. 
Magnons,  i.e., the quanta of spin-waves, carry magnetic momentum $\mu_{\rm{B}}$ and heat $k_{\rm{B}}$.
Those thermomagnetic transport properties are described by the WF law [Eq. (\ref{eqn:WFmay})].
The exchange interaction between the nearest neighbor spins in each FI (whose width is $L$) is $J $
and the magnetic field in the left (right) FI along the $z$-axis is denoted by $ B_{l(r)}$. 
The time reversal symmetry is broken by the ferromagnetic order and the magnetic field. 
The boundary spins ${\bf S}_{\Gamma_{l}}$ in the left FI and ${\bf S}_{\Gamma_{r}}$ in the right FI are relevant to the transport of such magnons
and they are weakly exchange-coupled with the strength $J_\textrm{ex}$ ($\ll  J$). 
\label{fig:order} }
\end{center}
\end{figure}

A very promising strategy to look for  low-dissipation spintronics devices
is to use pure magnon currents \cite{MagnonSpintronics} which are collective low-energy excitations created and transported in insulating magnets,
therefore in the complete absence
of any  conducting metallic elements \cite{magnon2,Trauzettel}.

It was indeed shown that the magnetic insulator Y$_3$Fe$_5$O$_{12}$ (YIG) can convert a heat flow into a spin voltage \cite{uchidainsulator} which was detected  by attaching a Pt thin film to YIG and using the inverse spin Hall effect \cite{ishe} to convert
spin-wave currents into electric currents \cite{spinwave}.
Such an observation has a natural interpretation in terms of the spin Seebeck effect (SSE) \cite{uchidametal}  for magnetic insulators.
In such materials, the SSE was shown \cite{xiao2010,adachi,hoffman2013} to be driven by magnons only which are the low-energy collective magnetic excitation of spins.
The Onsager reciprocal effect \cite{Onsager2}, the spin Peltier effect (SPE) corresponding to the generation of a magnon heat current by a spin current, has also been recently observed experimentally \cite{Peltier}.

In this paper, we analyze magnon and heat currents induced by magnetic field and temperature differences in insulating ferromagnetic  junctions as depicted in Fig. \ref{fig:order}.  We calculate 
 all transport coefficients for magnon transport and establish Onsager relations between them.  We predict some universal thermomagnetic relations for magnon transport in ferromagnetic junctions
 and establish a magnon analog of the Wiedemann-Franz (WF) law \cite{WFgermany} which is known to govern charge transport at low temperatures  \cite{AMermin,kittel}. With respect to the SSE or SPE, we demonstrate that the magnon Seebeck and Peltier coefficients behave universally in the low temperature limit. We also show that these features  are extremely robust with respect to multi-magnon effects.
However, we find that  magnon-magnon interactions can give rise to deviations from the Onsager reciprocity relation between the
magnon Seebeck and Peltier coefficients. Such a mechanism may offer an explanation for recent experiments by Dejene {\textit{et al.}} \cite{OnsagerExperiment} where deviations from the Onsager relation between the spin Seebeck and Peltier coefficients were observed.

The paper is organized as follows. In Sec.~II we introduce the model system for a ferromagnetic junction. In Sec.~III we describe the magnon transport in harmonic approximation driven by  gradients of magnetic fields or heat and derive all Onsager transport coefficients perturbatively.
There we also find the WF law and the universal behavior of the Seebeck and Peltier coefficients. In Sec.~IV
we discuss  multi-magnon effects such as three- and four magnon terms. In Sec. V we give some concrete estimates for YIG systems.
Finally, we summarize and give some conclusions in Sec.~VI. The technical details are deferred to App.~A-E.
\\
{\section{System}}
\label{system1}
We consider a magnetic junction formed by two ferromagnetic insulators (FI), as illustrated in Fig. \ref{fig:order}.
The temperature of the left (right) FI is $T_{l(r)}$ and the cross-section area of the junction interface is $\cal{A}$.
Due to a finite overlap of the wave functions  there exists in general a finite exchange interaction between the spins located at the boundaries between the two FIs.
Thus, only the boundary spins, denoted as  ${\bf S}_{\Gamma_{l}}$ and ${\bf S}_{\Gamma_{\rm{r}}}$ in the left and right FI, respectively (see Fig. \ref{fig:order}), are relevant for magnon transport across the junction interface.
The exchange interaction  between the two FIs may  be described by the Hamiltonian \cite{KKPD,KPD}
$ {\cal{H}}_{\rm{ex}}  = -J_{\rm{ex}} \sum_{\langle \Gamma_{l} \Gamma_{r} \rangle} {\bf S}_{\Gamma_{l}} \cdot {\bf S}_{\Gamma_{r}}$,
where $ J_{\rm{ex}} > 0$ is the  exchange interaction, weakly coupling the two FIs.
Assuming magnetic order along the magnetic field, defining the z-direction, we perform a
Holstein-Primakoff expansion \cite{HP,adachi} to leading order,
$S_{l/r}^+ = \sqrt{2S} a_{l/r}^{\dagger} +{\cal{O}}(S^{-1/2}) $ and
$S_{l/r}^z = S - a_{l/r}^\dagger a_{l/r}$,  where $ [a_{l}, a_{r}^{\dagger }]= \delta _{l, r} $,
we obtain
\begin{eqnarray}
 {\cal{H}}_{\rm{ex}} =  - J_{{\rm{ex}}} S \sum_{{\mathbf{k}}_{\perp }}  \sum_{k_x, k_x^{\prime}}
a_{\Gamma _{l},  {\mathbf{k}}}  a^{\dagger }_{\Gamma _{r}, {\mathbf{k}}^{\prime}} + {\rm{H. c.}},
\label{eqn:nonconservedScattering} 
\end{eqnarray}
where ${\mathbf{k}}=(k_x, k_y, k_z)$, ${\mathbf{k}}^{\prime}=(k_x^{\prime}, k_y, k_z)$, ${\mathbf{k}}_{\perp }=(0, k_y, k_z)  $,
and 
the bosonic operator $a_{\Gamma_{r/l}}^{\dagger}$ ($a_{\Gamma_{r/l}}$) creates (annihilates) a boundary magnon at the right/left FI. 
We note that the $k_x$-momentum of magnons is not conserved at the sharp junction interface, whereas the perpendicular momentum ${\mathbf{k}}_{\perp }$ is conserved. 
The tunneling Hamiltonian  $ {\cal{H}}_{\rm{ex}}$ thus gives the time-evolution of the magnon number operators in both FIs 
and generates the magnon and heat currents.
In obtaining Eq. (\ref{eqn:nonconservedScattering}), we have assumed large spins $S \gg 1$ \cite{uchidainsulator,adachi} 
and hence the ${\cal{O}}(S^{0})$ term in Eq. (\ref{eqn:nonconservedScattering}) indeed becomes negligible.

Assuming cubic lattices, each of the three-dimensional FIs can be described by a Heisenberg spin Hamiltonian in the presence of a Zeeman term.
The time reversal symmetry is broken by the assumed ferromagnetic order and by the magnetic field.
Within the long wave-length approximation and in the continuum limit, the magnon dispersion relation in each FI reads
\begin{equation}
 \omega _{{{ {k}}}}^{l(r)} = 2J S a^2 k^2 +g\mu_{\rm{B}}B_{l(r)}, 
 \label{eqn:Jdef}
 \end{equation}
where $J >0$ is the isotropic exchange interaction between the nearest neighbor spins in each FI,
$a $ denotes the lattice constant, $k \equiv  \mid {\bf {k}} \mid  $ is the wave vector modulus,
and $B_{l(r)} $ is the magnetic field for magnons in the left (right) FI along the $z$-axis.
We  assume that $ J_{\rm{ex}} \ll  J $  and treat $ {\cal{H}}_{\rm{ex}}$  perturbatively.
We remark that in addition to the 
tunneling Hamiltonian given by Eq. (\ref{eqn:nonconservedScattering}),
other ${\cal{O}}(J_{\rm{ex}}S)$-terms 
actually arise from  ${\cal{H}}_{\rm{ex}}$ due to the ferromagnetic order 
and act as effective magnetic fields for the boundary magnons.
Including such effects, an effective magnetic field 
in Eq. (\ref{eqn:Jdef}) is introduced.
Thus, even in the absence of an external magnetic field,   $B_{l(r)}  > 0$ for the boundary magnons.
\\

{\section{Onsager coefficients}}

The magnetic field and temperature differences defined by 
$\Delta B \equiv  B_{r} -  B_{l}$ and $   \Delta T \equiv  T_{r} -  T_{l}$, respectively,
generate the magnon and heat  currents  \cite{AMermin,mahan}  ${\cal{I}}_{\rm{m}} $ and ${\cal{I}}_{Q}$,
where \cite{sign}
${\cal{I}}_{\rm{m}} = - i  (J_{\rm{ex}}S/\hbar ) 
\sum_{\mathbf{k},{\mathbf{k}}^{\prime}} 
g \mu _{\rm{B}}  a_{\Gamma_{l},{\mathbf{k}}}  a^{\dagger }_{\Gamma_{r},{\mathbf{k}}^{\prime}} + {\rm{H. c.}}$ 
and
${\cal{I}}_{Q} = - i  (J_{\rm{ex}}S/\hbar ) 
\sum_{\mathbf{k},{\mathbf{k}}^{\prime}} 
\omega _{{{ {k}}}}^{l}  a_{\Gamma_{l},{\mathbf{k}}}  a^{\dagger }_{\Gamma_{r},{\mathbf{k}}^{\prime}} + {\rm{H. c.}}$ 
Within the linear response regime, each Onsager coefficient $L^{ij}$ ($i, j = 1, 2$) and the matrix $\hat{L}$ are defined by
\begin{eqnarray}
\begin{pmatrix}
\langle {\cal{I}}_{\rm{m}} \rangle  \\  \langle {\cal{I}}_{Q} \rangle
\end{pmatrix}
=
\begin{pmatrix}
L^{11} & L^{12} \\ L^{21} & L^{22}
\end{pmatrix}
\begin{pmatrix}
-     \Delta B  \\  \Delta T
\end{pmatrix}
\equiv \hat{L}
\begin{pmatrix}
- \Delta B  \\  \Delta T
\end{pmatrix}.
\label{eqn:LinearResponseIJ}
\end{eqnarray}
A straightforward perturbative calculation in $J_{\rm{ex}}$ \cite{Supple} based on the Schwinger-Keldysh \cite{rammer,tatara} formalism and up to ${\cal{O}} (J_{\rm{ex}}^2)$
 gives each coefficient
\begin{widetext}
\begin{subequations}
\begin{eqnarray}
{{ L}}^{11} &=& \frac{ (g \mu _{\rm{B}})^2  {\cal A}}{2 \hbar }  \Big(\frac{J_{\rm{ex}}}{4 \pi J  a}\Big)^2
\sum_{n=1}^{\infty } [- {\rm{Ei}}(-n \epsilon^2 )] {\rm{e}}^{-nb},    
  \label{eqn:GB23}  \\
{{ L}}^{12} &=&  \frac{ g \mu _{\rm{B}}k_{\rm{B}}  {\cal A}}{2 \hbar }   \Big(\frac{J_{\rm{ex}}}{4 \pi J a}\Big)^2     
 \Big[\sum_{n=1}^{\infty } (1/n+b)[- {\rm{Ei}}(-n \epsilon^2 )] {\rm{e}}^{-nb} +{\rm{Li}}_1({\rm{e}}^{-b})\Big],
   \label{eqn:GT23}    \\
{{L}}^{21} &=&    \frac{ g \mu _{\rm{B}}k_{\rm{B}} T  {\cal A}}{2 \hbar }  \Big(\frac{J_{\rm{ex}}}{4 \pi J a}\Big)^2     
 \Big[\sum_{n=1}^{\infty } (1/n+b)[- {\rm{Ei}}(-n \epsilon^2 )] {\rm{e}}^{-nb} +{\rm{Li}}_1({\rm{e}}^{-b})\Big], 
 \label{eqn:LB23}  \\
{{ L}}^{22} &=&  \frac{  k_{\rm{B}}^2 T  {\cal A}}{2 \hbar }    \Big(\frac{J_{\rm{ex}}}{4 \pi J a}\Big)^2
\Big[3{\rm{Li}}_2({\rm{e}}^{-b})+ 2 b{\rm{Li}}_1({\rm{e}}^{-b})  
+ \sum_{n=1}^{\infty } \Big( \frac{2}{n^2}  + \frac{2 b}{ n} + b^2 \Big)[- {\rm{Ei}}(-n \epsilon^2)] {\rm{e}}^{-nb} 
\Big],
 \label{eqn:LT23}
\end{eqnarray}
\end{subequations}
\end{widetext}
where 
\begin{eqnarray}
b \equiv   \frac{g \mu _{\rm{B}}B}{k_{\rm{B}} T}.
\label{eqn:b}
\end{eqnarray}
We have denoted $  B_{l}\equiv  B  $  and $T_{l} \equiv  T$ for simplicity,
and a phenomenological magnon lifetime \cite{demokritov}  $\tau $ in $  \epsilon \equiv  \hbar \beta /(2 \tau) \ll  1$,  
where $ \beta \equiv (k_{\rm{B}} T)^{-1}$,
has been introduced and is mainly due to nonmagnetic impurity scatterings.
Here, ${\rm{Li}}_{s}(z) = \sum_{n=1}^{\infty} z^n/n^s$ is the polylogarithm function and
 ${\rm{Ei}}(-n \epsilon^2) = \gamma + {\rm{ln}} \mid  n \epsilon^2  \mid + {\cal{O}}( n \epsilon^2)  $ the exponential integral, 
where $\gamma $ is the Euler constant.
The coefficients are seen to satisfy the Onsager relation  \cite{Onsager2,AMermin,mahan} 
\begin{eqnarray}
 L^{21}=   T\cdot  L^{12}.
  \label{eqn:OnsagerRelation}
\end{eqnarray}

\begin{figure}[h]
\begin{center}
\includegraphics[width=6cm,clip]{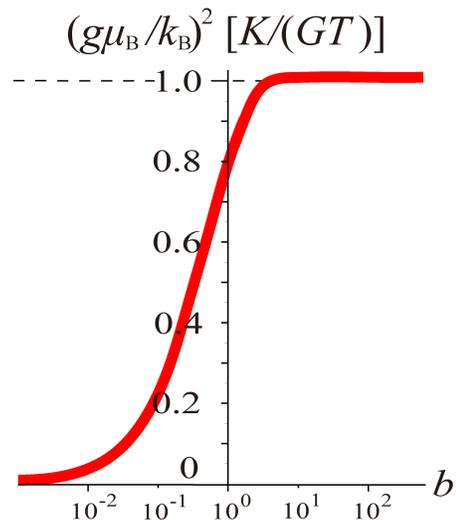}
\caption{(Color online)
Plot of  the ratio $ (g \mu _{\rm{B}}/k_{\rm{B}})^2[K/(GT)] $ as function of $b $ where $ \epsilon =10^{-10}$.
At low temperatures $b  = {\cal{O}}(10)$, the ratio reaches the constant `$1$' and the WF law  for magnon transport [Eq. (\ref{eqn:WFmay})] is realized.
\label{fig:WF_MagnonTransport}
}
\end{center}
\end{figure}

The coefficient $L^{11}$ is identified with the magnetic magnon conductance 
$G$, 
and, in analogy to charge transport \cite{AMermin,mahan,Supple}, 
the thermal magnon conductance $K$ is defined by  $K \equiv  L^{22} -  L^{21} L^{12}(L^{11})^{-1} $. 
From Eqs. (\ref{eqn:GB23})-(\ref{eqn:LT23}) we obtain  the  thermomagnetic ratio  $K/G$, characterizing magnon and heat transport.  
Its behavior is plotted in Fig. \ref{fig:WF_MagnonTransport}.
At low temperatures \cite{PhononWF} (i.e., $1\ll  b  $),
the ratio  becomes  linear in temperature (see also Fig. \ref{fig:WF_MagnonTransport}),
\begin{eqnarray}
 \frac{K}{G} \stackrel{\rightarrow }{=}   \Big(\frac{k_{\rm{B}}}{g \mu _{\rm{B}}}\Big)^2    T,  
   \        {\rm{when}}     \            \hbar /(2\tau )  \ll      k_{\rm{B}} T \ll  g \mu _{\rm{B}} B.
   \label{eqn:WFmay}
\end{eqnarray}
Therefore, in analogy to charge transport in metals \cite{WFgermany,AMermin,kittel}, 
we refer to this behavior as the {\it Wiedemann-Franz law for magnon transport}.
The constant $\cal{L}$ analogous to the Lorenz number becomes 
\begin{eqnarray}
 {\cal{L}} =   \Big(\frac{k_{\rm{B}}}{g \mu _{\rm{B}}}\Big)^2,
   \label{eqn:Lorentz}
\end{eqnarray}
where the role of the charge $e$ is played by $g \mu _{\rm{B}}$.
The magnetic Lorenz number is independent of any material parameters except the $g$-factor which
is material specific. Interestingly, the WF law holds in the same way for magnons, which are bosonic excitations,
as for electrons which are fermions. 
The linear-in-$T$ behavior can be traced back to the Onsager relation, Eq. (\ref{eqn:OnsagerRelation}), and is independent of microscopic details \cite{SupplYIG}.

Similarly, in analogy to charge transport in metals \cite{AMermin,mahan},
we refer to ${\cal{S}} \equiv  L^{12}/L^{11}$ as magnon Seebeck coefficient 
(i.e., thermomagnetic power),
and $\Pi  \equiv  L^{21}/L^{11}$ as  magnon Peltier coefficient.
The Onsager relation Eq. (\ref{eqn:OnsagerRelation}) provides the Thomson relation (i.e., Kelvin-Onsager relation \cite{spincal})
\begin{eqnarray}
  \Pi  = T {\cal{S}}.
  \label{eqn:Thomson}
\end{eqnarray}
At low temperatures, $   \hbar /(2\tau )  \ll    k_{\rm{B}} T \ll  g \mu _{\rm{B}} B$, the coefficients reduce to  \cite{Supple}
\begin{eqnarray}
 {\cal{S}} \stackrel{\rightarrow }{=} \frac{B}{T}, \,\,\,\,\,\,\,\,\,\,\,\,\,
 \Pi  \stackrel{\rightarrow }{=} B.
\label{eqn:Peltier}  
\end{eqnarray}
This is a remarkable result: 
The magnon Seebeck and Peltier coefficients become {\it universal} at low temperatures, i.e., 
they are completely independent of any material parameters (including the $g$-factor) and are solely determined by the applied magnetic field and temperature.

Finally, we remark that at low temperatures, $   \hbar /(2\tau )  \ll   k_{\rm{B}} T \ll  g \mu _{\rm{B}} B$, each Onsager coefficient ${{L}}^{ij}$ ($i, j =1, 2$) and the thermal magnon conductance behave in terms of $\epsilon $ as
\begin{eqnarray}
 {{L}}^{ij} \sim  {\rm{ln}}\epsilon, \,\,\,\,\,\,\,\,\,\,\,\,\,  K  \sim  {\rm{ln}}\epsilon.
 \label{eqn:Peltier2777}  
\end{eqnarray}
Thus, all coefficients show a weak logarithmic dependence on $\tau$, {\it i.e.}, ${{L}}^{ij}, K \sim {\rm{ln}}\tau$.
In addition, Eq. (\ref{eqn:LinearResponseIJ}) implies that  both currents arise from terms of order $ {\cal{O}} (J_{\rm{ex}}^2)$.
Therefore, even when an electric field is applied to the interface, the resulting Aharonov-Casher phase \cite{casher} cannot play any significant  role in the transport of such noncondensed magnons. 
Moreover, even when a magnetic field difference $ \Delta B \ne 0$ is generated, the noncondensed magnon current becomes essentially a dc one. 
This is in sharp contrast to the condensed magnon current \cite{KKPD} which arises from the $ {\cal{O}}(J_{\rm{ex}})$-term. 
\\

{\section{Multi-magnon effects}}
So far we have  considered the transport of essentially noninteracting magnons. 
Now we  take multi-magnon effects into account.
Two kinds of  effects are considered below; the first one corresponds to a three-magnon splitting \cite{kurebayashi} which arises due to  higher order terms in the $1/S$-expansion of the Holstein-Primakoff transformation \cite{HP}, while the second one appears because of magnon-magnon interactions \cite{KKPD,KPD,Kevin2} and is due to the anisotropy \cite{Supple} of the exchange interaction between neighboring spins in each FI.

We begin by considering higher order terms of the Holstein-Primakoff transformation \cite{HP},
$ S_i^+ =  \sqrt{2S} [1-a_i^\dagger a_i / (4S)] a_i +{\cal{O}}(S^{-3/2})  $, 
and thereby include the three-magnon splittings, $a_i^\dagger a_i a_i/S + {\rm{H. c.}}$,  into $ {\cal{H}}_{\rm{ex}} $.
One can expect that each coefficient becomes smaller because such terms correspond to the replacement of the operator $a_i$ with  $[1-a_i^\dagger a_i / (4S)] a_i $.
A straightforward calculation \cite{Supple} based again on the Schwinger-Keldysh \cite{rammer,tatara} formalism gives the simple result for the modified matrix $\hat{L}_2$ up to ${\cal{O}} (J_{\rm{ex}}^2)$
\begin{equation}
\hat{L}_2= \Big[1- \frac{1}{4 \pi^{3/2}S} \Big(\frac{k_{\rm{B}}T}{2JS}\Big)^{3/2} {\rm{Li}}_{3/2}({\rm{e}}^{-b })   \Big] \hat{L}.
\label{eqn:three}
\end{equation}
Eq. (\ref{eqn:three}) shows that, although each coefficient becomes smaller due to the three-magnon splittings, 
the modified matrix $\hat{L}_2$ is still characterized by the noninteracting one $\hat{L}$ [Eq. (\ref{eqn:LinearResponseIJ})]; 
$\hat{L}_2\propto \hat{L} $.
Consequently, the  Onsager and the Thomson relations, Eqs. (\ref{eqn:OnsagerRelation}) and (\ref{eqn:Thomson}), remain satisfied. 
In addition, the magnon WF law [Eq. (\ref{eqn:WFmay})] and the Seebeck and Peltier coefficients, 
[Eqs. (\ref{eqn:Peltier})],
at low temperature remain valid.
These thermomagnetic properties are therefore robust against the three-magnon splittings.
One can show \cite{Supple} that these properties actually
 hold in any order of the  $1/S$-expansion of the Holstein-Primakoff
  transformation. 

An anisotropic \cite{KKPD,KPD,Kevin2} exchange interaction between  nearest-neighbor spins in each FI gives rise to magnon-magnon interactions of the type \cite{Supple}
$  {\cal{H}}_{\rm{m}} = - J_{\rm{m}}   \sum_{\langle i j\rangle}  a_{i}^{\dagger } a_{j}^{\dagger }  a_{i} a_{j} $. 
The symbol $\langle i j\rangle$ indicates summation over nearest-neighbor spins in each FI. 
The magnitude and the sign of the interaction $J_{\rm{m}}$ depends on the anisotropy \cite{Supple} of the exchange interaction. We assume here 
a small anisotropy $ \mid  J_{\rm{m}}   \mid   \    \ll   J $. 
A straightforward but tedious calculation \cite{Supple} 
gives us the modified matrix
up to ${\cal{O}} (J_{\rm{ex}}^2 J_{\rm{m}})$
\begin{equation}
 \hat{L}_3 =
\begin{pmatrix}
  L^{11} + \delta L^{11}  &   L^{12}  \\ 
  L^{21} + \delta L^{21} &  L^{22} 
\end{pmatrix},
\label{eqn:magnonmagnon}
\end{equation}
where
\begin{subequations}
\begin{eqnarray}
\delta  L^{11} & = &   \frac{ J_{\rm{ex}}^2 J_{\rm{m}}  \sqrt{k_{\rm{B}}T} \tau  {\cal{A}}}{16 \sqrt{2S} \pi ^{5/2} \hbar^2  J^{5/2} }   
                                          (g \mu _{\rm{B}}\Lambda)^2   {\rm{Li}}_{3/2}({\rm{e}}^{-b })  \nonumber    \\
& \times &   {\rm{ln}}(\sqrt{2JS \beta } a \Lambda /\epsilon ),    
\label{eqn:deltaG_B}   \\
\delta  L^{21}  & = &    \frac{ J_{\rm{ex}}^2J_{\rm{m}}  \sqrt{k_{\rm{B}}T} \tau {\cal{A}}}{16 \sqrt{2S} \pi ^{5/2} \hbar^2  J^{5/2}} 
   g \mu _{\rm{B}}  \Lambda ^2  {\rm{Li}}_{3/2}({\rm{e}}^{-b })    \nonumber    \\
      & \times &     \Big[  J S (a\Lambda)^2 -  \epsilon ^2/(2 \beta)  + [ J S (a\Lambda)^2 +  g\mu_{\rm{B}}  B] \nonumber    \\
                    & \times & {\rm{ln}}(\sqrt{2JS \beta } a \Lambda /\epsilon ) \Big],   
                     \label{eqn:deltaL_B}   
\end{eqnarray}
\end{subequations}
with $ a \Lambda  \equiv  \sqrt{5/(JS\beta)} $.
These results imply the violation of the Onsager relation in Eq. (\ref{eqn:OnsagerRelation}) [and of the Thomson relation in Eq. (\ref{eqn:Thomson})]
due to  
magnon-magnon interactions.
Notice that $\delta  L^{11}  =  {\cal{O}}(J_{\rm{ex}}^2J_{\rm{m}})$ and $\delta  L^{21}  =  {\cal{O}}(J_{\rm{ex}}^2J_{\rm{m}})$, while $  L^{11}  =  {\cal{O}}(J_{\rm{ex}}^2J_{\rm{m}}^0)$ and $  L^{21}  =  {\cal{O}}(J_{\rm{ex}}^2J_{\rm{m}}^0)$.
We recall that the Onsager reciprocal relation \cite{Onsager2,Casimir,OnsagerRelation} could in principle be already violated in the noninteracting case since the time reversal symmetry is broken by the ferromagnetic order and the magnetic field right from the outset. 
Still, we have microscopically found that the relation remains satisfied
even in the presence of the three-magnon and higher order splitting terms of the Holstein-Primakoff transformation \cite{HP}.
However, the anisotropy induced magnon-magnon interaction $J_{\rm{m}} $ provides a `nonlinearity' $\delta  L^{21} =  {\cal{O}}(J_{\rm{ex}}^2J_{\rm{m}})$  in terms of the perturbative terms ($J_{\rm{ex}}$ and $J_{\rm{m}}$),
and consequently the matrix $\hat{L}_3 $ cannot be reduced to the form $\hat{L}=  {\cal{O}}(J_{\rm{ex}}^2J_{\rm{m}}^0)$.
Using a mean field argument or a   more rigorous microscopic calculation \cite{Supple}, one can show that  the magnon-magnon interaction 
acts as an effective magnetic field and that the total magnetic field difference $ \Delta B_{\rm{tot}} $ may be written as
$ \Delta B_{\rm{tot}} = (1+b_{\rm{m}})\Delta B  $ with $ b_{\rm{m}}={\cal{O}}(J_{\rm{m}})$.
The term $ b_{\rm{m}}$ gives $\delta  L^{11}={\cal{O}}(J_{\rm{m}})$ and $\delta  L^{21}={\cal{O}}(J_{\rm{m}})$.
The Onsager relation is thus violated due to the nonlinearity caused by the anisotropy induced magnon-magnon interaction.
The magnitude of the effective magnetic field difference can be estimated by 
$  b_{\rm{m}} \sim \delta L^{21}/L^{21}$.

These multi-magnon contributions,  $\delta L^{11}$ and $\delta  L^{21}$, 
generally affect also the thermomagnetic properties and can lead to deviations from our previous results.
However,  at low temperatures, where the WF law, Eq. (\ref{eqn:WFmay}), and the universality of Seebeck and Peltier coefficients hold,   these deviations become negligible \cite{Supple} because  
$|\delta L^{ij}/L^{ij}|\ll 1$
for typical  parameter values (see below).

So far we have assumed bulk FIs (see Fig. \ref{fig:order}) where
magnetic dipole-dipole interactions  are negligible \cite{tupitsyn}. Such dipolar effects, however, become important in thin films, resulting in a modified
dispersion for magnons \cite{demokritov}. Still,  the WF law remains valid in this case too \cite{SupplYIG}, underlining the universality of this law.
\\

{\section{Estimates for experiments}}
The magnon currents can be experimentally measured by using, for instance, the method proposed in Refs. [\onlinecite{magnon2,dipole,KKPD}]. 
Since the magnons, being moving magnetic dipoles with  magnetic moment $g\mu_{\rm{B}}  {\bf e}_z$, produce electric fields,  magnon currents can be detected by measuring the resulting voltage drop perpendicular to the current direction and magnetic field. 
For an estimate, we assume the following experimental parameter values \cite{uchidainsulator,adachi,demokritov,MagnonPhonon,YIGdw,Tarucha}: 
$ J =  100 $meV, $J_{\rm{ex}}=10 $meV, $J_{\rm{m}}=1 $meV, $ a  = 1 $\AA,  ${\cal{A}} = 3$cm$^2$,  
$g =2$, $\tau =100 $ns, $\Delta T = 0.5$K, and $ B  = 50$mT ($5$T)  and $T = 300 $K ($0.7$K) for the high (low)\cite{PhononWF} temperature regime.
Using a similar set-up as in  Ref. [\onlinecite{KKPD}], we find that the resulting voltage drop is in the mV ($\mu $V) range for high (low) temperatures.
Although small, such values are within experimental reach and are actually  about $ 10^6$ ($10^3$) times larger than the one ($\sim $nV) predicted for 
 currents in condensed magnon systems \cite{KKPD}. 
Alternatively, attaching a metal (e.g. Pt) to the FIs and using the inverse spin Hall effect \cite{ishe} to convert magnon currents into electric currents \cite{spinwave},
the magnon currents could also be detected \cite{uchidainsulator,uchidametal,Peltier} by measuring the resulting Hall voltage in the metal.
Finally, we mention that the temperature difference $\Delta  T$ can be experimentally  produced by applying  microwaves of different frequencies to each FI \cite{ultrahot} or by local laser heating \cite{MagnonPhonon,MagnonSupercurrent}.
\\

{\section{Summary}}
We have studied the thermomagnetic transport behavior of a ferromagnetic insulating junction and determined the Onsager coefficients in linear response regime. We found 
that at low temperatures the magnon transport obeys an analog of the Wiedemann-Franz law where the ratio of heat to magnon conductance is linear in temperature.  
Like its electronic counterpart the WF law found here  is $\it{universal}$ and does not depend on material parameters except the $g$-factor. Quite remarkably, 
it exhibits the same  linear-in-$T$ behavior at low temperatures as the one for electronic transport in spite of 
the fact that the quantum-statistical properties of bosons and fermions are fundamentally different, in particular in the low temperature regime where quantum effects dominate.

The temperature scale, however, for electrons is given by the Fermi temperature ($\sim 10^4$ K for normal metals), while it is the magnetic field for magnons (a few Kelvins). Obviously, these two scales
are very different, and that is why the electronic counterpart of the WF law is valid typically at much higher temperatures than the magnonic one, but in both cases the WF law applies when the system temperature is low {\it relative} to its respective temperature scale.
Moreover, we  showed that
the magnon Seebeck and Peltier coefficients become universal at low temperatures.

As an outlook we mention that it would be interesting to explore the regime beyond the weak junction coupling studied here and
see if the WF law can be extended to such a regime as well.

Finally, it would be interesting to test our predictions experimentally in candidate systems like insulating ferromagnets
such a YIG material in the bulk or thin film limit.
\\

\begin{acknowledgments}
We thank S. E. Nigg, R. Tiwari,  K. van Hoogdalem, and A. Zyuzin for stimulating discussions. 
We acknowledge support by the Swiss NSF, the NCCR QSIT ETHZ-Basel, 
by the JSPS Postdoctoral Fellow for Research Abroad (No. 26-143), JSPS Research Fellow (No. 25-2747), 
the Japanese-Swiss Science and Technology Programme supported by JSPS and ETHZ (KN), 
and by ANR under Contract No. DYMESYS (ANR 2011-IS04-001-01) (PS). 
\end{acknowledgments}


\appendix

\section{MAGNON AND HEAT CURRENTS}
\label{sec:current1}

Assume cubic lattices, each three-dimensional FI  is described by a Heisenberg spin Hamiltonian in the presence of a Zeeman term,
\begin{eqnarray}
    {\cal{H}}_{\rm{FI}} =  - \sum_{\langle i j\rangle} {\bf S}_i \cdot {\bf J} \cdot {\bf S}_j -  g\mu_{\rm{B}}  {\bf B}_{l(r)} \cdot \sum_i {\bf S}_i,
\label{eqn:Heisenberg}
\end{eqnarray}
where ${\bf J}$ denotes a diagonal $3\times 3$-matrix with $\textrm{diag}({\bf J}) = J\{ 1, 1,\eta\}$. 
The exchange interaction between neighboring spins in the ferromagnetic insulator is $J >0$,
$  \eta $ denotes the anisotropy of the spin Hamiltonian,
and ${\bf B}_{l(r)} = B_{l(r)} {\bf e}_z$ is an applied magnetic field to the left (right)  FI (${\bf e}_z$ denotes the unit vector along the $z$-axis).
The symbol $\langle i j\rangle$ indicates summation over neighboring spins in each FI and ${\bf S}_i$ denotes the spin of length $S$ at lattice site $i$.
Within our microscopic calculation \cite{KPD}, we find that in the continuum limit, 
the magnon-magnon interaction $  {\cal{H}}_{\rm{m}}$ could arise from the $\eta \not=1$ anisotropic spin Hamiltonian
 ${\cal{H}}_{\rm{FI}} $ as the ${\cal{O}}(1-\eta)$ term 
\begin{eqnarray}
  {\cal{H}}_{\rm{m}}  
=  - J_{\rm{m}} a^3  \int  d{\mathbf{r}} \   a^\dagger (\mathbf{r})  a^\dagger (\mathbf{r})  a (\mathbf{r})  a(\mathbf{r}),
\label{eqn:MagnonMagnonint}
\end{eqnarray}
where $ J_{\rm{m}} \equiv  - J (1-\eta ) = {\cal{O}}(S^0)$, the lattice constant $a$, and 
$ [a(\mathbf{r}), a^\dagger ({\mathbf{r}}^{\prime})]=  \delta ({\mathbf{r}} - {\mathbf{r}}^{\prime})$.
Therefore the magnon-magnon interaction does not influence the magnon transport between $ \eta =1 $ isotropic FIs in any significant manner \cite{KKPD,KPD} 
and we can neglect them in the isotropic case. 
Assuming the isotropic FI,  within the long wave-length approximation, the magnon dispersion  in each FI 
is given by Eq. (\ref{eqn:Jdef}) in the main text, i.e.,
$ \omega _k^{l(r)} = 2J S a^2 k^2 +g\mu_{\rm{B}}B_{l(r)}$.

We then consider a magnetic junction formed by two ferromagnetic insulators, as illustrated in Fig. \ref{fig:order}.
The temperature of the left (right) FI is $T_{l(r)}$ and the cross-section area of the junction interface is $\cal{A}$.
Due to a finite overlap of the wave functions,  there exists in general a finite exchange interaction between the spins located at the boundaries between the two FIs.
Thus, only the boundary spins, denoted as  ${\bf S}_{\Gamma_{l}}$ and ${\bf S}_{\Gamma_{r}}$ in the left and right FI, respectively (see Fig. \ref{fig:order}), are relevant for magnon transport across the junction interface.
The exchange interaction  between the two FIs may  be described by the Hamiltonian \cite{KKPD,KPD}
$ {\cal{H}}_{\rm{ex}}  = -J_{\rm{ex}} \sum_{\langle \Gamma_{l} \Gamma_{r} \rangle} {\bf S}_{\Gamma_{l}} \cdot {\bf S}_{\Gamma_{r}}$,
where $ J_{\rm{ex}} > 0$ is the  exchange interaction, weakly coupling the two FIs.
Assuming magnetic order along the magnetic field, defining the z-direction, we perform a
Holstein-Primakoff expansion \cite{HP,adachi} to leading order,
$S_{ l/r}^+ = \sqrt{2S} a_{l/r}^{\dagger} +{\cal{O}}(S^{-1/2}) $ and
$S_{ l/r}^z = S - a_{l/r}^\dagger a_{l/r}$,  where $ [a_{l}, a_{r}^{\dagger }]= \delta _{l, r} $,
we obtain Eq. (\ref{eqn:nonconservedScattering}) in the main text.

The tunneling Hamiltonian  $ {\cal{H}}_{\rm{ex}}$  gives the time-evolution of the magnon number operators in the FIs and generates the magnon and  heat currents \cite{AMermin,mahan} in the junction. 
The Heisenberg equation of motion provides
the magnon current operator $ {\cal{I}}_{\rm{m}} $ and  heat current  operator ${\cal{I}}_{Q}$ by  \cite{sign}
${\cal{I}}_{\rm{m}} = - i  (J_{\rm{ex}}S/\hbar ) 
\sum_{\mathbf{k}} \sum_{{\mathbf{k}}^{\prime}}g \mu _{\rm{B}}  a_{\Gamma_{l},{\mathbf{k}}}  a^{\dagger }_{\Gamma_{r},{\mathbf{k}}^{\prime}} + {\rm{H. c.}}$ 
and
${\cal{I}}_{Q} = - i  (J_{\rm{ex}}S/\hbar ) 
\sum_{\mathbf{k}} \sum_{{\mathbf{k}}^{\prime}}\omega _k^{l}  a_{\Gamma_{l},{\mathbf{k}}}  a^{\dagger }_{\Gamma_{r},{\mathbf{k}}^{\prime}} + {\rm{H. c.}}$ 
Using the Schwinger-Keldysh \cite{rammer,tatara} formalism and treating $ {\cal{H}}_{\rm{ex}}$ perturbatively ($J_{\rm{ex}}\ll J$), 
they can be evaluated  up to ${\cal{O}} (J_{\rm{ex}}^2)$ 
\begin{subequations}
\begin{eqnarray}
     \langle {\cal{I}}_{\rm{m}} \rangle  
  &=&   \frac{(J_{\rm{ex}}S)^2 a^2}{2 \pi  L^2}  \sum_{{\mathbf{k}}, k_x^{\prime}}  \int d \omega   g \mu _{\rm{B}}     \label{eqn:definition} \\  
  &\times  &  ({\cal{G}}^{\rm{<}}_{l, {\mathbf{k}}, \omega } {\cal{G}}^{\rm{>}}_{r, {\mathbf{k^{\prime}}}, \omega }
              -  {\cal{G}}^{\rm{>}}_{l, {\mathbf{k}}, \omega }  {\cal{G}}^{\rm{<}}_{r, {\mathbf{k^{\prime}}}, \omega }  
)  , 
\nonumber 
\\
     \langle {\cal{I}}_{Q} \rangle  
  &=&    \frac{(J_{\rm{ex}}S)^2 a^2}{2 \pi  L^2}  \sum_{{\mathbf{k}}, k_x^{\prime}}  \int d \omega  
 \omega _k^{l}  \label{eqn:definitionJ}  \\  
  &\times  &  ({\cal{G}}^{\rm{<}}_{l, {\mathbf{k}}, \omega } {\cal{G}}^{\rm{>}}_{r, {\mathbf{k^{\prime}}}, \omega }
              -  {\cal{G}}^{\rm{>}}_{l, {\mathbf{k}}, \omega }  {\cal{G}}^{\rm{<}}_{r, {\mathbf{k^{\prime}}}, \omega }  
)  , 
\nonumber 
\end{eqnarray}
\end{subequations}
where $L$ is the width of the FI (Fig. \ref{fig:order}) and ${\cal{ G}}^{\rm{< (>)}}$ are the bosonic lesser (greater) Green functions.
The both currents arise from the $ {\cal{O}} (J_{\rm{ex}}^2)$-terms; 
$\langle {\cal{I}}_{\rm{m}} \rangle = {\cal{O}}(J_{\rm{ex}}^2)$ and $\langle {\cal{I}}_{Q} \rangle = {\cal{O}}(J_{\rm{ex}}^2)$.

We phenomenologically \cite{tatara} introduce a life-time $\tau $ of magnons mainly due to nonmagnetic impurity scatterings  into Green functions (e.g., the retarded Green function $ {\cal{G}}^{\rm{r}}_{{\mathbf{k}}, \omega }=[\hbar \omega - \omega _{\mathbf{k}}+i \hbar /(2 \tau )]^{-1} $), and regard it as a constant  \cite{demokritov}. This gives
\begin{eqnarray}
{\cal{G}}^{\rm{<}}_{l, {\mathbf{k}}, \omega } {\cal{G}}^{\rm{>}}_{r, {\mathbf{k^{\prime}}}, \omega }
                -  {\cal{G}}^{\rm{>}}_{l, {\mathbf{k}}, \omega } {\cal{G}}^{\rm{<}}_{r, {\mathbf{k^{\prime}}}, \omega } 
                              &=& - \Big(\frac{\hbar }{\tau }\Big)^2   
                                      \mid  {\cal{G}}^{\rm{r}}_{l, {\mathbf{k}}, \omega }   \mid^2 
                                      \mid  {\cal{G}}^{ \rm{r}}_{r, {\mathbf{k^{\prime}}}, \omega }   \mid^2 \nonumber  \\
                              &\times &  [n (\omega _{\mathbf{k}}^{l}) -  n (\omega _{\mathbf{k^{\prime}}}^{r})],
\label{eqn:BoseDistributionGreen}
\end{eqnarray}
where the Bose-distribution function $ n (\omega _{\mathbf{k}}) =  (e^{\beta \omega _{\mathbf{k}}} -1)^{-1} $ and $\beta \equiv  1/(k_{\rm{B}}T)$. 
Thus the currents become
\begin{widetext}
\begin{subequations}
\begin{eqnarray}
 \langle {\cal{I}}_{\rm{m}}  \rangle 
&=& -  \frac{2 }{\hbar }\Big(\frac{J_{{\rm{ex}}} S a}{L}\Big)^2
 \sum_{{\mathbf{k}}}  g \mu_{\rm{B}}
 \sum_{k_x^{\prime}} 
  [n(\omega _{\mathbf{k}}^{l}) -n(\omega _{\mathbf{k^{\prime}}}^{r})]
 \frac{\hbar /(2\tau ) }{[2JS a^2({k_x^{\prime}}^2 - k_x^2)]^2+[\hbar /(2\tau )]^2},     \label{eqn:goalllll}   \\
   \langle {\cal{I}}_{\rm{Q}}  \rangle 
&=& -  \frac{2 }{\hbar }\Big(\frac{J_{{\rm{ex}}} S a}{L}\Big)^2
 \sum_{{\mathbf{k}}}  \omega _{\mathbf{k}}^{l}  
 \sum_{k_x^{\prime}} 
  [n(\omega _{\mathbf{k}}^{l}) -n(\omega _{\mathbf{k^{\prime}}}^{r})]
 \frac{\hbar /(2\tau )}{[2JS a^2({k_x^{\prime}}^2 - k_x^2)]^2+[\hbar /(2\tau )]^2}.   \label{eqn:goal2llll} 
\end{eqnarray}
\end{subequations}
For large $L$, we can replace the sums by integrals.
The summation over $k_x^{\prime}$ in Eqs. (\ref{eqn:goalllll}) and (\ref{eqn:goal2llll}) thus becomes
\begin{subequations}
\begin{eqnarray}
 \sum_{k_x^{\prime}}
 [n(\omega _{\mathbf{k}}^{l}) -n(\omega _{\mathbf{k^{\prime}}}^{r})]  
 \frac{\hbar /(2\tau )}{[2JS a^2({k_x^{\prime}}^2 - k_x^2)]^2+[\hbar /(2\tau )]^2}   
 &\stackrel{\rightarrow }{=}&  \frac{L}{2 \pi}\sqrt{ \frac{\beta }{2JS a^2}} \int d x^{\prime} 
[n(\omega _{\mathbf{k}}^{l}) -n(\omega _{\mathbf{k^{\prime}}}^{r})]
\frac{\epsilon }{({x^{\prime}}^2-x^2)^2+\epsilon^2 }  \\
  & \stackrel{\rightarrow }{=} & \frac{L}{2 \pi}\sqrt{ \frac{\beta }{2JS a^2}}
 \int d x^{\prime} [n(\omega _{\mathbf{k}}^{l}) -n(\omega _{\mathbf{k^{\prime}}}^{r})] \pi \delta ({x^{\prime}}^2-x^2), \label{eqn:delta2} 
\end{eqnarray}
\end{subequations}
\end{widetext}
where $   x^2  \equiv 2JS a^2 \beta k_x^2$,   $   {x^{\prime}}^2  \equiv 2JS a^2 \beta {k_x^{\prime}}^2$, and $ \epsilon \equiv  \hbar \beta /(2 \tau)$.
In the last equation [Eq. (\ref{eqn:delta2})], we have assumed that $\tau$ is  large such that $  \epsilon \ll 1 $.

We see that both currents [Eqs. (\ref{eqn:goalllll}) and (\ref{eqn:goal2llll})] are characterized by the difference  of  Bose-distribution functions 
$n (\omega)$. 
After integration over $x^{\prime}$, Eq. (\ref{eqn:delta2})  reduces to $  n (\omega _{\mathbf{k}}^{l}) -  n (\omega _{\mathbf{k}}^{r}) $.
Within the linear response regime, this difference becomes 
\begin{eqnarray}
    n (\omega _{\mathbf{k}}^{l}) -  n (\omega _{\mathbf{k}}^{r})   \approx 
\begin{cases}
\beta  g \mu_{\rm{B}} \frac{e^{\beta \omega _{\mathbf{k}}}}{(e^{\beta \omega _{\mathbf{k}}}-1)^2}  \Delta  B,  
& {\textrm{ for }}   \Delta T  = 0,
  \\ 
 - \frac{\beta  \omega _{\mathbf{k}}}{T}  \frac{e^{\beta \omega _{\mathbf{k}}}}{(e^{\beta \omega _{\mathbf{k}}}-1)^2}  \Delta  T,
& {\textrm{ for }} \Delta B  = 0.
\label{eqn:BoseDistributionGreen2}
\end{cases}
\end{eqnarray}
We have introduced the magnetic field and temperature differences 
\begin{eqnarray}
 \Delta B \equiv  B_{r} -  B_{l}, \   \Delta T \equiv  T_{r} -  T_{l},
\end{eqnarray}  
and have denoted $  B_{l}\equiv  B  $  and $T_{l} \equiv  T$ for convenience. 
The difference of the Bose-distribution functions has been expanded in powers of $\Delta T\ll T$ and $\Delta B\ll B$. 
Eq. (\ref{eqn:BoseDistributionGreen2}) yields
\begin{eqnarray}
 \frac{[n (\omega _{\mathbf{k}}^{l}) -  n (\omega _{\mathbf{k}}^{r})] \mid _{\Delta B=0}/\Delta T}
{[n (\omega _{\mathbf{k}}^{l}) -  n (\omega _{\mathbf{k}}^{r})] \mid _{\Delta T=0}/\Delta B}
 = - \frac{\omega _{\mathbf{k}}^{l}}{g \mu_{\rm{B}}T}.
\label{eqn:BoseDistributionGreen3}
\end{eqnarray}
Within linear response, we can also approximate the magnetic field terms in all Green functions in Eq.~(\ref{eqn:BoseDistributionGreen}) by $B$.
Finally, one can see that Eq. (\ref{eqn:BoseDistributionGreen3}) is responsible for the Onsager reciprocal relation, given in Eq. (\ref{eqn:OnsagerRelation}) in the main text.

Thus, we see that the magnon and heat currents are characterized by the difference  of the Bose-distribution functions, and
the difference gives the linear responses [Eq. (\ref{eqn:BoseDistributionGreen2}) and (\ref{eqn:BoseDistributionGreen3})], 
and the Onsager relation, Eq. (\ref{eqn:OnsagerRelation}), holds accordingly.
This remains in place even when three-magnon splittings [Eq. (\ref{eqn:three222})] and any higher order terms of the Holstein-Primakoff expansion are taken into account
[see Eqs. (\ref{eqn:three-magnonsExpansion}) and (\ref{eqn:three-magnonsExpansionLLL}) as an example]:
 in any order of  the expansion,  the currents are characterized by the difference of the Bose-distribution functions [Eq. (\ref{eqn:BoseDistributionGreen2})]. 
Consequently, the Onsager relation [Eqs. (\ref{eqn:BoseDistributionGreen3}) and (\ref{eqn:OnsagerRelation})] holds.

This is changed by the anisotropy induced magnon-magnon interaction and the Onsager relation becomes violated, as we shall see below.
\\

\section{ONSAGER COEFFICIENTS}
\label{sec:current12}

Within the linear response regime [Eq. (\ref{eqn:BoseDistributionGreen2})],
Eqs. (\ref{eqn:goalllll}) and (\ref{eqn:goal2llll}) provide the Onsager coefficients $L^{ij}$ in Eqs. (\ref{eqn:GB23})-(\ref{eqn:LT23}) of the main text. 
The coefficients, $L^{21}$ and $L^{12}$,  are seen to satisfy the Onsager relation
Eq. (\ref{eqn:OnsagerRelation}). 
The coefficient $L^{11}$ is identified with the magnetic magnon conductance \cite{magnon2} $G$, 
and the thermal magnon conductance $K$ is defined by  $K \equiv  L^{22} -  L^{21} L^{12}/L^{11} $. In analogy to charge transport \cite{AMermin,mahan}, this definition follows from the condition that the magnon current $\langle {\cal{I}}_{\rm{m}} \rangle $ induced by the applied thermal difference $\Delta T$ be zero. This gives rise to an {\it induced} magnetic field difference 
$\Delta B_{\rm{ind}}=\Delta T L^{12}/L^{11}$,  and thus to the thermal magnon current $\langle {\cal{I}}_{Q} \rangle=K \Delta T$ (in the absence of an applied field gradient, i.e., $\Delta B=0$).

Eqs. (\ref{eqn:GB23})-(\ref{eqn:LT23}) determine  the general thermomagnetic ratio  $K/G$ for magnon and heat transport, see Fig. \ref{fig:WF_MagnonTransport} in the main text.
At low temperatures,  $1\ll  b \equiv  g \mu _{\rm{B}}B/(k_{\rm{B}} T)$,
the ratio reduces to Eq. (\ref{eqn:WFmay}) in the main text (see also Fig. \ref{fig:WF_MagnonTransport}).
We note that the temperature should be still such that $\epsilon \ll 1$ remains satisfied, i.e., $  \hbar /(2 \tau)  \ll k_{\rm{B}} T \ll  g \mu _{\rm{B}} B $.
In the opposite limit, $\epsilon \gg 1$, both currents, Eqs. (\ref{eqn:goalllll}) and  (\ref{eqn:goal2llll}), are seen to be exponentially vanishing 
as $T \rightarrow 0$.

\begin{figure}[h]
\begin{center}
\includegraphics[width=7cm,clip]{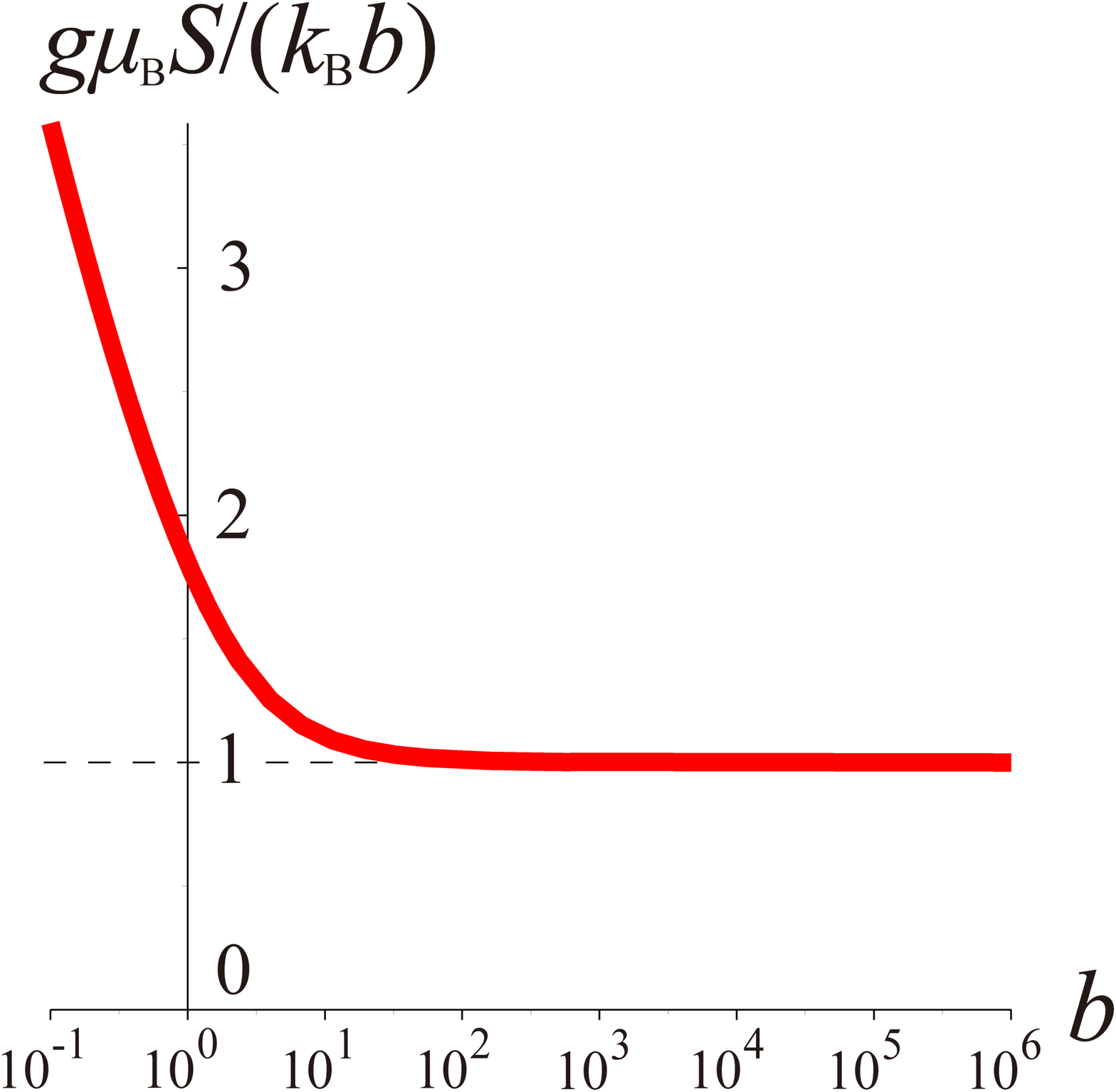}
\caption{(Color online)
Plot of magnon Seebeck coefficient $\cal{S}$  as function of $b $ where $ \epsilon =10^{-10}$.
At low temperatures $b  = {\cal{O}}(10^2)$, 
the rescaled coefficient $g \mu _{\rm{B}} {\cal{S}}/(k_{\rm{B}}b) $ approaches unity asymptotically and realizes the universal relation [Eq. (\ref{eqn:Peltier})].
\label{fig:Seebeck}
}
\end{center}
\end{figure}

The Onsager relation [Eq. (\ref{eqn:OnsagerRelation})] provides the Thomson relation (i.e., the Kelvin-Onsager relation \cite{spincal}) in Eq. (\ref{eqn:Thomson}) of the main text.
At low temperatures,   $\hbar /(2\tau )  \ll       k_{\rm{B}} T \ll  g \mu _{\rm{B}} B$,
the magnon Seebeck and Peltier coefficients reduce to Eq. (\ref{eqn:Peltier}) in the main text (see Fig. \ref{fig:Seebeck}), and, quite strikingly, become universal.

We remark that at low temperatures, each Onsager coefficient ${{L}}^{ij}$ and the thermal magnon conductance behave in terms of $\epsilon $ as
in Eq. (\ref{eqn:Peltier2777}) in the main text.
Thus, all coefficients show a weak logarithmic dependence on $\tau$, i.e., ${{L}}^{ij}, K \sim {\rm{ln}}\tau$.

\section{YIG THIN FILM}
\label{sec:YIG}

So far we have assumed
that both FIs are of bulk shape such that surface effects due to magnetic dipole-dipole interactions are negligible \cite{tupitsyn}.
However, it is interesting to consider also a thin film geometry since such systems are of great experimental interest, in particular for YIG films \cite{demokritov}.

We show now that the WF law also holds for a thin film, where now the length of the junction in x-direction L  (see Fig. \ref{fig:order}) is short compared to all other dimensions.
Due to the dipole-dipole interaction, the magnon dispersion relation in each FI changes and  becomes \cite{demokritov} 
$ \omega _{{{ {k}}}}^{l(r)} = D (k^2-k_{\rm{m}}^2)^2 +g\mu_{\rm{B}}B_{l(r)} $, where $k_{\rm{m}}\sim  10^4$/cm for e.g. YIG thin films \cite{demokritov}.
The parameter $D$ is due to  the long-range dipole-dipole interaction as well as the exchange interaction between the nearest-neighbor spins.
The main contribution comes from $k \approx   k_{\rm{m}}$ where 
$ \omega _{{{ {k}}}}^{l(r)} \approx  4D k_{\rm{m}}^2 (k-k_{\rm{m}})^2 +g\mu_{\rm{B}}B_{l(r)} $ for $k \sim    k_{\rm{m}}$.
As we have seen in the main text, the WF law is realized in strong magnetic fields $B_{l(r)}\sim $ a few T (leading to a large energy gap). 
Repeating the same perturbative calculation as before, but under the restriction that only the lowest magnon subband in $x$-direction ($ k_x = k_x^{\prime} =  0  $)  needs to be taken into account
due to finite-size quantization (which actually simplifies the calculation),
the Onsager coefficients in thin films become
\begin{widetext}
\begin{subequations}
\begin{eqnarray}
{{ L}}^{11} &=& \frac{ \tau  {\cal A}}{ 4 \pi \hbar^2 D k_{\rm{m}}^2}  \Big(\frac{J_{\rm{ex}}S a}{L}\Big)^2
(g \mu _{\rm{B}})^2 
{\rm{Li}}_{0}({\rm{e}}^{-b}),
  \label{eqn:GB23YIG}  \\
{{ L}}^{12} &=&  \frac{ \tau  {\cal A}}{ 4 \pi \hbar^2 D k_{\rm{m}}^2}  \Big(\frac{J_{\rm{ex}}S a}{L}\Big)^2
 g \mu _{\rm{B}} k_{\rm{B}}
\Big[{\rm{Li}}_{1}({\rm{e}}^{-b})+ b{\rm{Li}}_{0}({\rm{e}}^{-b})\Big],
   \label{eqn:GT23YIG}    \\
{{L}}^{21} &=&    \frac{ \tau  {\cal A}}{ 4 \pi \hbar^2 D k_{\rm{m}}^2}  \Big(\frac{J_{\rm{ex}}S a}{L}\Big)^2
g \mu _{\rm{B}} k_{\rm{B}}  T
\Big[{\rm{Li}}_{1}({\rm{e}}^{-b})+ b{\rm{Li}}_{0}({\rm{e}}^{-b})\Big],
 \label{eqn:LB23YIG}  \\
{{ L}}^{22} &=&   \frac{ \tau  {\cal A}}{ 4 \pi \hbar^2 D k_{\rm{m}}^2}  \Big(\frac{J_{\rm{ex}}S a}{L}\Big)^2
k_{\rm{B}}^2 T
\Big[2 {\rm{Li}}_{2}({\rm{e}}^{-b})
+2 b {\rm{Li}}_{1}({\rm{e}}^{-b})
+ b^2 {\rm{Li}}_{0}({\rm{e}}^{-b})\Big].
 \label{eqn:LT23YIG} 
\end{eqnarray}
\end{subequations}
\end{widetext}
The Onsager relation, ${{L}}^{21}  = T  \cdot  {{L}}^{12} $, again holds.

In particular, the ratio $K/G$ reduces to\cite{PhononWF}
\begin{eqnarray}
 \frac{K}{G} \stackrel{\rightarrow }{=}   \Big(\frac{k_{\rm{B}}}{g \mu _{\rm{B}}}\Big)^2    T,  \,\,\,\,
     {\rm{when}}     \             k_{\rm{B}} T \ll  g \mu _{\rm{B}} B. 
   \label{eqn:WF_YIG}
\end{eqnarray}
Again, the ratio is linear in temperature, and, quite remarkably, again with a universal prefactor (in particular independent of the dipole-dipole interaction $D$).
Thus, we see that the
WF law holds also for thin films and in the presence of dipole-dipole interactions. 
Thus, we conclude that the linear-in-$T$ behavior is extremely robust against microscopic details.

We note that each Onsager coefficient itself [Eqs. (\ref{eqn:GB23YIG})-(\ref{eqn:LT23YIG})] drastically changes in terms of  temperature dependence as well as  magnon lifetime  $\it{etc.}$ as compared to the bulk results (see main text). However, these differences all cancel out in the ratio $K/G$ and the linear-in-temperature behavior remains.

Finally, at low temperatures $ k_{\rm{B}} T \ll  g \mu _{\rm{B}} B$, 
the magnon Seebeck and Peltier coefficients for  thin films are reduced to 
${\cal{S}} \stackrel{\rightarrow }{=} {B}/{T}$ and
$ \Pi  \stackrel{\rightarrow }{=} B$
as for the bulk case given in the main text.
\\

\section{MAGNON-MAGNON INTERACTIONS}
\label{sec:NLeffectAppendix}
\label{sec:current15}

Assuming an anisotropic \cite{KKPD,KPD,Kevin2} exchange interaction among the nearest neighboring spins in each FI [Eq. (\ref{eqn:Heisenberg})],  
magnon-magnon interactions in Eq. (\ref{eqn:MagnonMagnonint})
may arise from such an anisotropic Heisenberg spin Hamiltonian.
The sign and the magnitude of the magnon-magnon interaction $J_{\rm{m}}^{l(r)}$ depends on the anisotropy in the left (right) FI
and it is assumed to be small $ \mid  J_{\rm{m}}^{l(r)}   \mid   \    \ll   J $ to be treated perturbatively; 
in Eqs. (\ref{eqn:deltaG_B})  and (\ref{eqn:deltaL_B}), they are set $J_{\rm{m}}^{l}=J_{\rm{m}}^{r}\equiv J_{\rm{m}} $ for simplicity.

Using the Schwinger-Keldysh formalism \cite{rammer,tatara}, we find up to ${\cal{O}} (J_{\rm{ex}}^2 J_{\rm{m}})$
\begin{eqnarray}
\langle {\cal{I}}_{\rm{m}}  \rangle
  &= &  \frac{(J_{\rm{ex}}S)^2  J_{\rm{m}}^{l} a^2 }{2 \pi^{5/2} L^2}   
                    \Big(\frac{k_{\rm{B}}T}{2JS}\Big)^{3/2}  {\rm{Li}}_{3/2}({\rm{e}}^{-b })
                     \sum_{{\mathbf{k}}, k_x^{\prime}}
\int   d \omega    g \mu _{\rm{B}}   \nonumber  \\
  &\times & {\rm{Re}} 
         ({\cal{G}}_{{l},{\mathbf{k}},\omega }^{\rm{r}}
          {\cal{G}}_{{l},{\mathbf{k}},\omega }^{\rm{r}}
          {\cal{G}}_{{r},{\mathbf{k^{\prime}}},\omega }^{\rm{r}}
      + {\cal{G}}_{{l},{\mathbf{k}},\omega }^{\rm{r}}
          {\cal{G}}_{{l},{\mathbf{k}},\omega }^{\rm{r}}
          {\cal{G}}_{{r},{\mathbf{k^{\prime}}},\omega }^{\rm{<}}   \nonumber   \\
   &+&{\cal{G}}_{{l},{\mathbf{k}},\omega }^{\rm{<}}
          {\cal{G}}_{{l},{\mathbf{k}},\omega }^{\rm{r}} 
          {\cal{G}}_{{r},{\mathbf{k^{\prime}}},\omega }^{\rm{a}} 
     +   {\cal{G}}_{{l},{\mathbf{k}},\omega }^{\rm{<}}
          {\cal{G}}_{{l},{\mathbf{k}},\omega }^{\rm{a}}
          {\cal{G}}_{{r},{\mathbf{k^{\prime}}},\omega }^{\rm{a}} 
          )    \nonumber   \\
&-& ({l} \leftrightarrow {r}) ,
 \label{eqn:ThermalMagnon_MagMag}
\end{eqnarray}
where ${\cal{G}}^{\rm{a(r/</>)}}$ is the bosonic advanced (retarded/lesser/greater) Green function. 
Thus the magnon current $\langle {\cal{I}}_{\rm{m}} \rangle = {\cal{O}}(J_{\rm{m}}^{l(r)})$ arises  
as the nonlinearity in terms of the perturbative terms ($J_{\rm{m}}^{l(r)}$ and $J_{\rm{ex}}$).
After some straightforward manipulations [Eq. (\ref{eqn:delta2})],
Eq. (\ref{eqn:ThermalMagnon_MagMag}) is reduced to the form 
\begin{eqnarray}
\langle {\cal{I}}_{\rm{m}}  \rangle &= & 
\Big[ {\cal{C}}_1 (J_{\rm{m}}^{l} {\cal{N}}^{l}+J_{\rm{m}}^{r}{\cal{N}}^{r})   
\label{eqn:J_m-expansion}    \\
&+&\int   dk  \  {\cal{C}}_2 (J_{\rm{m}}^{l} {\cal{N}}^{l}-J_{\rm{m}}^{r}{\cal{N}}^{r})
[n (\omega _{\mathbf{k}}^{l}) -  n (\omega _{\mathbf{k}}^{r})]\Big] \Delta B,   \nonumber 
\end{eqnarray}
where ${\cal{C}}_{1(2)}$ is a $\Delta B$- and $\Delta T$-independent constant, and $ {\cal{N}}^{l(r)}$ defined by
\begin{subequations}
\begin{eqnarray}
 {\cal{N}}^{l} & \equiv &   \frac{\sqrt{\pi}}{4 a ^3} \Big(\frac{k_{\rm{B}}T}{2JS}\Big)^{3/2}  
{\rm{Li}}_{3/2}({\rm{e}}^{-b }),  \label{eqn:FBL}  \\
  {\cal{N}}^{r} &=&   {\cal{N}}^{l} + {\cal{O}}(\Delta B) +  {\cal{O}}(\Delta T), \label{eqn:FBR} 
\end{eqnarray}
\end{subequations}
is obtained by integrating the Bose-distribution function $n (\omega _{\mathbf{k}}^{l(r)})$ over the wavevector.
Eq. (\ref{eqn:J_m-expansion}) indicates that the perturbation of the magnon-magnon interaction $J_{\rm{m}}^{l(r)}$ work as the magnetic field difference $\Delta B$ expansion.
A mean field analysis actually shows that the magnon-magnon interaction works as an effective magnetic field.
The ${\cal{C}}_{1}$-term arises from 
$ {\cal{G}}_{{l},{\mathbf{k}},\omega }^{\rm{r}} {\cal{G}}_{{l},{\mathbf{k}},\omega }^{\rm{r}} {\cal{G}}_{{r},{\mathbf{k^{\prime}}},\omega }^{\rm{r}}$ in Eq. (\ref{eqn:ThermalMagnon_MagMag})
and the ${\cal{C}}_{2}$-term from the rest.
Since the difference of the Bose-distribution functions $n (\omega _{\mathbf{k}}^{l}) -  n (\omega _{\mathbf{k}}^{r})$ indeed gives   the responses to $\Delta B$ and $\Delta T$ by itself [Eq. (\ref{eqn:BoseDistributionGreen2})], the ${\cal{C}}_{2}$-term becomes irrelevant  in the linear response regime; 
the response to the temperature difference $\Delta T$ actually arises also in the magnon-magnon interactions (i.e., from the ${\cal{C}}_{2}$-term), but it is accompanied with the magnetic field difference  as such $\Delta B \cdot \Delta T$.
Therefore any linear responses to the temperature difference in the magnon-magnon interactions
cannot be found in the corresponding magnon current [Eq. (\ref{eqn:J_m-expansion})] or 
the modified matrix $\hat{L}_3$ [Eq. (\ref{eqn:magnonmagnon})].
We remark that the same remarks apply to the difference of magnon-magnon interactions $\Delta  J_{\rm{m}}\equiv   J_{\rm{m}}^{r}-J_{\rm{m}}^{l}$.
The response to $\Delta  J_{\rm{m}}$ actually arises from the ${\cal{C}}_{2}$-term 
[Eqs. (\ref{eqn:J_m-expansion}), (\ref{eqn:FBR}), and (\ref{eqn:BoseDistributionGreen2})], 
but it is accompanied with  nonlinear responses such as  $\Delta B \cdot \Delta T$ and $(\Delta B)^2 $.
Therefore, any responses to $\Delta  J_{\rm{m}}$ cannot affect the modified matrix $\hat{L}_3$ [Eq. (\ref{eqn:magnonmagnon})].

Finally, focusing on the linear response regime, the magnon current due to the magnon-magnon interactions becomes
\begin{eqnarray}
\langle {\cal{I}}_{\rm{m}}  \rangle = {\cal{C}}_1 (J_{\rm{m}}^{l} {\cal{N}}^{l}+J_{\rm{m}}^{r} {\cal{N}}^{r})   \Delta B.
\label{eqn:J_m-expansion2}
\end{eqnarray}
This gives $\delta  L^{11}$ in Eq. (\ref{eqn:deltaG_B}),  where they are set  $J_{\rm{m}}^{l}=J_{\rm{m}}^{r} \equiv J_{\rm{m}}$.
Using $\delta  L^{11}$, the constant is given by 
${\cal{C}}_1 = -  \delta L^{11}/(J_{\rm{m}}^{l} {\cal{N}}^{l}+J_{\rm{m}}^{r}{\cal{N}}^{r})$.

The main mechanism behind the generation of the coefficient $\delta L^{21}$ due to the magnon-magnon interactions remains in place. 
The heat current $\langle {\cal{I}}_{Q}  \rangle ={\cal{O}}(J_{\rm{m}}^{l(r)})$ can be evaluated in the same way
and it is indeed expressed in the same form as Eq. (\ref{eqn:ThermalMagnon_MagMag}) in terms of the Green functions  
up to ${\cal{O}} (J_{\rm{ex}}^2 J_{\rm{m}})$
\begin{eqnarray}
\langle {\cal{I}}_{Q}   \rangle
  &=&  \frac{(J_{\rm{ex}}S)^2  J_{\rm{m}}^{l} a^2}{2 \pi^{5/2}L^2}   
                    \Big(\frac{k_{\rm{B}}T}{2JS}\Big)^{3/2}  {\rm{Li}}_{3/2}({\rm{e}}^{-b })
                     \sum_{{\mathbf{k}}, k_x^{\prime}}
\int   d \omega   \omega _k^{l}   \nonumber  \\
  &\times & {\rm{Re}}  
         ({\cal{G}}_{{l},{\mathbf{k}},\omega }^{\rm{r}}
          {\cal{G}}_{{l},{\mathbf{k}},\omega }^{\rm{r}}
          {\cal{G}}_{{r},{\mathbf{k^{\prime}}},\omega }^{\rm{r}}
      + {\cal{G}}_{{l},{\mathbf{k}},\omega }^{\rm{r}}
          {\cal{G}}_{{l},{\mathbf{k}},\omega }^{\rm{r}}
          {\cal{G}}_{{r},{\mathbf{k^{\prime}}},\omega }^{\rm{<}}   \nonumber   \\
   &+&{\cal{G}}_{{l},{\mathbf{k}},\omega }^{\rm{<}}
          {\cal{G}}_{{l},{\mathbf{k}},\omega }^{\rm{r}} 
          {\cal{G}}_{{r},{\mathbf{k^{\prime}}},\omega }^{\rm{a}} 
     +   {\cal{G}}_{{l},{\mathbf{k}},\omega }^{\rm{<}}
          {\cal{G}}_{{l},{\mathbf{k}},\omega }^{\rm{a}}
          {\cal{G}}_{{r},{\mathbf{k^{\prime}}},\omega }^{\rm{a}} 
          )  \nonumber   \\
&-& ({l} \leftrightarrow {r}) .
 \label{eqn:ThermalMagnon_MagMagJ}
\end{eqnarray}
After some straightforward manipulations with the help of Eq. (\ref{eqn:delta2}), it  finally reduces to 
\begin{eqnarray}
\langle  {\cal{I}}_{Q}  \rangle &= &   \Big[ {\cal{C}}_1^{\prime} (J_{\rm{m}}^{l} {\cal{N}}^{l}+J_{\rm{m}}^{r} {\cal{N}}^{r})   
  \label{eqn:J_m-expansion3}   \\
&+&\int   dk  \  {\cal{C}}_2^{\prime} (J_{\rm{m}}^{l} {\cal{N}}^{l}-J_{\rm{m}}^{r} {\cal{N}}^{r})
[n (\omega _{\mathbf{k}}^{l}) -  n (\omega _{\mathbf{k}}^{r})]\Big] \Delta B,   \nonumber 
\end{eqnarray}
where ${\cal{C}}_{1(2)}^{\prime} $ is the corresponding $\Delta B$- and $\Delta T$-independent constant.
This result has the same qualitative form as the magnon current Eq. (\ref{eqn:J_m-expansion}).
Within linear response, it becomes  [Eq. (\ref{eqn:BoseDistributionGreen2})]
\begin{eqnarray}
\langle  {\cal{I}}_{Q}  \rangle =  {\cal{C}}_1^{\prime} (J_{\rm{m}}^{l} {\cal{N}}^{l}+J_{\rm{m}}^{r} {\cal{N}}^{r})   \Delta B.
\label{eqn:J_m-expansion4}
\end{eqnarray}
This gives $ \delta L^{21}$ in Eq. (\ref{eqn:deltaL_B}), where  $J_{\rm{m}}^{l}=J_{\rm{m}}^{r} \equiv J_{\rm{m}}$.
The Onsager relation is violated by $ \delta L^{21}$ due to the magnon-magnon interactions.
Using $ \delta L^{21}$, the constant is given by 
${\cal{C}}_1^{\prime} = -  \delta L^{21}/(J_{\rm{m}}^{l} {\cal{N}}^{l}+J_{\rm{m}}^{r} {\cal{N}}^{r})$.

These results can be understood as follows;
the magnon-magnon interaction $J_{\rm{m}}^{l(r)}$ acts as an effective magnetic field,
and consequently the total magnetic field difference $ \Delta B_{\rm{tot}} $ may be written as 
$ \Delta B_{\rm{tot}} = (1+b_{\rm{m}})\Delta B  $,  
where $ b_{\rm{m}}={\cal{O}}(J_{\rm{m}}^{l}  {\cal{N}}^{l}+J_{\rm{m}}^{r} {\cal{N}}^{r})$ is the contribution of such magnon-magnon interactions.
This contribution $ b_{\rm{m}}$ gives rise to the ${\cal{C}}_1^{(\prime)}$-term 
[Eqs. (\ref{eqn:J_m-expansion2}) and (\ref{eqn:J_m-expansion4}), and  Eq. (\ref{eqn:FBL})], and leads to
$\delta  L^{11}={\cal{O}}(J_{\rm{m}}^{l} {\cal{N}}^{l}+J_{\rm{m}}^{r} {\cal{N}}^{r})$ and 
$\delta  L^{21}={\cal{O}}(J_{\rm{m}}^{l} {\cal{N}}^{l}+J_{\rm{m}}^{r} {\cal{N}}^{r})$.
Thus, the Onsager relation is  violated by the magnon-magnon interactions.
The magnitude of the effective magnetic field difference can be estimated by 
$  b_{\rm{m}} \sim \delta L^{21}/L^{21}$.

The contributions arising from magnon-magnon interactions,  $\delta L^{11}$ and $\delta  L^{21}$, 
generally affect also the thermomagnetic properties of magnon and heat transport.
However, at low temperatures, $  \hbar /(2 \tau)  \ll k_{\rm{B}} T \ll  g \mu _{\rm{B}} B $, 
the ratios reduce to $\delta L^{11}/L^{11} \sim  (3J_{\rm m}\tau/2\hbar) (k_{\rm B}T/SJ)^{3/2}$
and similarly for $\delta L^{21}/L^{21}$.
For typical  parameter values (see main text) we find that e.g. $\delta L^{11}/L^{11}\sim 0.1$. 
Therefore, $\hat{L}_3   \stackrel{\rightarrow }{=} \hat{L}  $ at low temperatures.
Thus, the Onsager and the Thomson relations, Eqs. (\ref{eqn:OnsagerRelation}) and (\ref{eqn:Thomson}), 
the WF law for magnon transport, Eq. (\ref{eqn:WFmay}), 
and the thermomagnetic properties of magnon Seebeck and Peltier coefficients, Eq. (\ref{eqn:Peltier}),
still hold at low temperatures.
\\

\section{THREE-MAGNON SPLITTINGS}
\label{sec:threeAppendix}
\label{sec:current16}

Until now, we have essentially used the linearized \cite{adachi} Holstein-Primakoff transformation 
$ S_i^+ = \sqrt{2S}[1-a_i^\dagger a_i / (2S)]^{1/2} a_i =   \sqrt{2S} a_i  + {\cal{O}}(S)^{-1/2}$.
Now, we consider the higher term ${\cal{O}}(S)^{-1/2}$, $ S_i^+ =  \sqrt{2S} [1-a_i^\dagger a_i / (4S)] a_i + {\cal{O}}(S)^{-3/2} $, 
and include the three-magnon splittings, $a_i^\dagger a_i a_i/S + {\rm{H. c.}}$, into the Hamiltonian ${\cal{H}}_{\rm{ex}}$ to be treated perturbatively.
We remark that when $S\rightarrow \infty $ (i.e. in the classical limit), the three-magnon splittings cease to work.
In that sense, the three-magnon splitting can be regarded as a quantum effect \cite{Peskin}.

Following the same procedure as before with the approximation Eq. (\ref{eqn:BoseDistributionGreen2}), 
the magnon current in linear response regime becomes in leading order 
\begin{eqnarray}
\langle  {\cal{I}}_{\rm{m}}  \rangle  &=&  \Big[1- \frac{1}{4 \pi^{3/2}S} \Big(\frac{k_{\rm{B}}T}{2JS}\Big)^{3/2}  {\rm{Li}}_{3/2}({\rm{e}}^{-b })\Big] \nonumber \\
&\times &
\begin{pmatrix}
L^{11} & L^{12} 
\end{pmatrix}
\begin{pmatrix}
-     \Delta B  \\  \Delta T 
\end{pmatrix} ,
 \label{eqn:three-magnonsExpansion}
\end{eqnarray}
which is of order 
$ {\cal{O}}(J_{\rm{ex}}^2)$.
The heat current can be evaluated in the same way and becomes  up to ${\cal{O}} (J_{\rm{ex}}^2)$
\begin{eqnarray}
\langle {\cal{I}}_{Q}  \rangle  &=&  \Big[1- \frac{1}{4 \pi^{3/2}S} \Big(\frac{k_{\rm{B}}T}{2JS}\Big)^{3/2} {\rm{Li}}_{3/2}({\rm{e}}^{-b })  \Big] 
\nonumber \\
&\times &
\begin{pmatrix}
L^{21} & L^{22} 
\end{pmatrix}
\begin{pmatrix}
-     \Delta B  \\  \Delta T
\end{pmatrix}   .
 \label{eqn:three-magnonsExpansionLLL}
\end{eqnarray}
Thus, the modified matrix $\hat{L}_2$ is given by [Eq. (\ref{eqn:LinearResponseIJ})]
\begin{equation}
\hat{L}_2= \Big[1- \frac{1}{4 \pi^{3/2}S} \Big(\frac{k_{\rm{B}}T}{2JS}\Big)^{3/2}   
{\rm{Li}}_{3/2}({\rm{e}}^{-b })   \Big] 
\hat{L}.
\label{eqn:three222}
\end{equation}
This means 
\begin{subequations}
\begin{eqnarray}
\hat{L}_2 &\propto &  \hat{L}    \label{eqn:three22299990000}    \\
                       &\stackrel{\rightarrow }{=}& \hat{L},       \     \     \       \    \      {\rm{when}}  \    \     T \rightarrow 0.
\label{eqn:three2229999}
\end{eqnarray}
\end{subequations}
Therefore, the Onsager and  the Thomson relations [Eqs. (\ref{eqn:OnsagerRelation}) and (\ref{eqn:Thomson})]  hold.
In addition, the WF law for magnon transport [Eq. (\ref{eqn:WFmay})] and 
the thermomagnetic properties [Eq. (\ref{eqn:Peltier})]
at low temperatures remain valid even in the presence of three-magnon splittings.

\bibliography{PumpingRef}

\begin{thebibliography}{45}
\bibitem[{sig()}]{sign}
\bibinfo{note}{The currents flow from the right FI to the left one when the
  sign of the currents is positive.}
\bibitem[{Pho()}]{PhononWF}
\bibinfo{note}{At low temperatures $T\sim 10^{-1}$K, phonon contributions are
  negligible \cite{adachiphonon}.}
\end{thebibliography}

\end{document}